\documentclass[11pt]{article}

\usepackage{makeidx}  
\makeindex
\usepackage{hyperref}

\usepackage{euscript}

\usepackage{mathrsfs}

\usepackage{latexsym}
\usepackage{amsmath, amssymb,graphics}
\usepackage[all]{xy}
\usepackage{color}

\usepackage{tikz}
\input xy
\xyoption{all}
\usepackage{euscript}
\usepackage{multirow}
\usepackage{etex, pictexwd,dcpic}
\usetikzlibrary{positioning}
\usetikzlibrary{shapes.geometric}
\usetikzlibrary{shapes.misc}
\usetikzlibrary{calc}
\usetikzlibrary{positioning}

\usepackage{pgflibraryarrows}
\usepackage{pgflibrarysnakes}

\usetikzlibrary{trees} 
\usetikzlibrary[trees] 

\usepackage{fullpage}

\newtheorem{theorem}{Theorem}[section]
\newtheorem{definition}[theorem]{Definition}
\newtheorem{lemma}[theorem]{Lemma}

\newtheorem{proposition}[theorem]{Proposition}

\newtheorem{remark}{Remark}

\numberwithin{equation}{section}

\renewcommand{\(}{\begin{equation*}}
\renewcommand{\)}{\end{equation*}}
\newcommand{\bea}{\begin{eqnarray*}}
\newcommand{\eea}{\end{eqnarray*}}
\newcommand{\R}{{\mathbb R}}
\newcommand{\C}{{\mathbb C}}
\newcommand{\Z}{{\mathbb Z}}
\newcommand{\Q}{{\mathbb Q}}

\def\proof {{Proof.}\hspace{7pt}}

\def\endofproof {\hfill{$\Box$}\\}

\newcommand{\beq}{\begin{equation}}
\newcommand{\eeq}{\end{equation}}

\numberwithin{equation}{section}

\renewcommand{\(}{\begin{equation}}
\renewcommand{\)}{\end{equation}}

\newcommand{\CC}{{\mathbb C}}

\newcommand{\CP}{\CC \text{P}}

\def\R{{\mathbb R}}
\def\Z{{\mathbb Z}}
\def\Q{{\mathbb Q}}
\def\C{{\mathbb C}}

\def\1{{\bf 1}}

\def\<{\langle}
\def\>{\rangle}

\numberwithin{equation}{section}


\newcommand{\U}{{\rm U}}

\begin{document}

\title{
String structures associated to indefinite Lie groups}

 \author{
  Hisham Sati$^1$ and Hyung-bo Shim$^2$\\
  }

\date{%
\footnotesize
    $^1$Division of Science and Mathematics, New York University Abu Dhabi,  Saadiyat Island, Abu Dhabi, UAE\\%
    $^2$Department of Mathematics, University of Pittsburgh, Pittsburgh, PA 15260, USA  \\[2ex]%
    \today
}

\maketitle

\begin{abstract} 
String structures have played an important role in algebraic topology, via elliptic 
genera and elliptic cohomology, in differential geometry, via the study of higher geometric structures,
and in physics, via partition functions. 
We extend the description of String structures from 
connected covers of the definite-signature orthogonal group ${\rm O}(n)$ to the indefinite-signature
orthogonal group ${\rm O}(p, q)$, i.e. from the Riemannian to the pseudo-Riemannian setting. 
This requires that we work at the unstable level, which makes
the discussion more subtle than the stable case. Similar, but much simpler, constructions hold for 
other noncompact Lie groups such as the unitary group ${\rm U}(p, q)$ and the symplectic group 
${\rm Sp}(p, q)$. 
This extension provides a starting point for 
an abundance of constructions in (higher) geometry and applications in physics. 
  
\end{abstract}


\tableofcontents

\newpage
\section{Introduction}

Lie groups play an important role in characterizing symmetries. Picking the appropriate Lie group 
allows for certain structures to be unambiguously defined. For example, a Riemannian structure on 
a manifold requires the principal frame bundle corresponding to the tangent bundle 
to have an orthogonal group as a structure group. To talk about orientations one needs 
the special orthogonal group, and to properly discuss spinors one needs to lift to the double cover,
which is the Spin group. All of these structures  are low degree phenomena 
which can be encoded uniformly and succinctly 
via the Whitehead tower of the orthogonal group (see \cite{SSS2} \cite{SSS3} \cite{9brane}). 

\medskip
In algebraic topology, one usually studies the Whitehead tower of the stable orthogonal 
group (see \cite{SSS2}). In particular, killing the third homotopy group leads to 
the stable String group (see \cite{St} \cite{ST}). There are many constructions and 
applications associated with String structures and to the String group in various areas of 
mathematics and physics. The following is a sample which is necessarily incomplete. 
In algebraic topology, String structures play a role of orientation for elliptic cohomology
 \cite{AHS} \cite{St} \cite{ST}. 
In differential geometry, String connections play a role in geometrically describing 
 bundles with the String group as a structure group 
 \cite{SSS3} \cite{Red} \cite{Wal} \cite{Bun} \cite{FSS}. In mathematical physics, 
 conditions for having String structures 
 arise as anomaly cancellation conditions \cite{Kil} \cite{SSS2} \cite{SSS3}.

\medskip
This paper aims to initiate a new angle on the subject.
We will be interested in the case of the indefinite orthogonal group ${\rm O}(p, q)$, which from a geometric 
point of view can be viewed as the structure group of the tangent bundle of a pseudo-Riemannian manifold 
of dimension $n = p+ q$. As in the Riemannian case, one is interested in considering oriented and then Spin 
pseudo-Riemannain 
manifolds and for that one needs to lift ${\rm O}(p, q)$ to appropriate groups ${\rm SO}(p, q)$ 
and ${\rm Spin}(p,q)$, respectively. 
This involves a lot of subtleties and unlike the Riemannian case, here it is 
 a priori not obvious which homotopy (sub-)groups to kill in order to get to the appropriate 
covering group. We spend some time discussing this before embarking on considering the corresponding 
indefinite String groups. In geometry and physics, the cases $p=1,2$ are particularly interesting, 
as these correspond to the Lorentz group and the conformal group, respectively. We address all 
cases, and for the most part there is a natural split into cases $p=1$, $p=2$ and $p\geq 3$.

\medskip
Another subtle matter in our discussion is that we need to work in the unstable range for the 
orthogonal groups, that is with ${\rm O}(n)$ for finite $n$, i.e. without taking a limit  on the rank 
as usually is done in the literature. For the Riemannian case, the Whitehead tower (cf. \cite{SSS2}) would 
be replaced by one involving finite $n$
\(
\xymatrix{
&&B{\rm String}(n)  \ar[d] &\\
&& B{\rm Spin}(n) \ar[d] \ar[r]^-{\tfrac{1}{2}p_1} & K(\Z,4)\\
&& B{\rm SO}(n) \ar[d] \ar[r]^-{w_2} & K(\Z/2,2)\\
X\ar[rr] \ar[urr] \ar[uurr] \ar[uuurr] && B{\rm O}(n) \ar[r]^-{w_1}& K(\Z/2,1)\;.
}
\)
where $w_1 \in  H^1(B{\rm O}(n);\Z/2)$ and $w_2 \in  H^2(B{\rm SO}(n);\Z/2)$ are the universal
first and second Stiefel-Whitney classes, while $\tfrac{1}{2}p_1 \in H^4(B{\rm Spin}(n);\Z)$ is the 
universal first Spin characteristic class. 

\medskip
This gives a homotopy theoretical construction of an 8-connected cover $(B{\rm O}(n))\langle 8\rangle$ 
of the classifying space of ${\rm O}(n)$. Having a classifying space points to presence of a group. 
Indeed, Stolz \cite{St} constructs a topological group $\widehat{G}$ 
 as an extension
\begin{equation}\label{Stolz}
1\longrightarrow 
{\rm Gauge}(P)\longrightarrow \widehat{G}\longrightarrow {\rm Spin}(n)
\longrightarrow 1\;,
\end{equation}
where $P\to {\rm Spin}(n)$ is the $P\U(\mathcal{H})$-principal bundle with $\mathcal{H}$ an 
infinite-dimensional separable Hilbert space  such that $B\widehat{G}$ is the 8-connected cover of 
$B{\rm O}(n)$. This topological group $\widehat{G}$ is homotopy equivalent to  the String group 
${\rm String}(n)$, defined via the fibration $K(\Z, 2) \to {\rm String}(n) \to {\rm Spin}(n)$, 
 so that  ${\rm String}(n)$ has a group structure.
This also has a differentiable structure \cite{NSW}. Since then many models of the String group have 
appeared, each having different desirable features (see e.g. the appendix of \cite{FSS} for seven such models).
For instance, in one model \cite{NSW}, ${\rm String}(n)$ is constructed as an extension as in
 eq. \eqref{Stolz} of Lie groups so that $\widehat{G}$ has a Fr\'echet-Lie group structure uniquely determined 
 up to isomorphism. 
The presence of such structures allows us to talk about ${\rm String}(n)$-principal bundles and String manifolds, 
regarding ${\rm String}(n)$ as a structure group.

\medskip
Studying the Spin group and its classifying space from the cohomology point of view requires 
understanding of the first generator, i.e., the generator in degree three of the cohomology of the  
Spin group or the generator in degree four of the corresponding classifying space
(see \cite{Wal2} \cite{FSS} for interesting relations between the two). It is known from \cite{Th} 
that the cohomology ring of $B{\rm Spin}$ in the stable case is generated by the Spin characteristic classes,
the degree four generator of which is $\tfrac{1}{2}p_1$. However, we are interested in the unstable case,
and indeed  it was shown by McLaughlin \cite{Mc} that 
$\frac{1}{2}p_1$ is also the generator of the cohomology $H^4(B{\rm Spin}(n);\Z)$.
The lifting of the structure group from ${\rm Spin}(n)$ to ${\rm String}(n)$ 
of a bundle over a manifold $X$ is then possible when the obstruction $\frac{1}{2}p_1(X)$ vanishes.
We will be interested in generalizing this result to the case of $B{\rm Spin}(p,q)$.

 
 \medskip
 String structures are interesting from the geometric point of view due to the relation between 
 the Riemannian geometry of a manifold and characteristic classes associated with String 
 structures on that manifold. 
The Stolz-H\"ohn conjecture says the following: Let $X$ be a smooth closed string manifold of 
dimension $4k$. If $X$ admits a Riemannian metric with positive Ricci curvature, then the 
Witten genus $\phi_W(X)$ vanishes.

\medskip
Homotopy-theoretically, String structures amount to having a String orientation, which in relation to 
modular forms gives a corresponding orientation to the spectrum of topological modular forms (tmf), 
$M{\rm String}\to {\rm tmf}$ \cite{AHS}. Conjecturally, the Witten genus is constructed as an index of 
a Dirac operator on loop space \cite{Wit}. This is the String analog to the theorem on vanishing 
$\hat{A}$-genus by Lichnerowicz \cite{Li} for Spin manifolds: Let $X$ be a smooth closed spin manifold 
of dimension $2k$. If $X$ admits a Riemannian affine connection with nonnegative and not identically 
zero Riemannian curvature, then the $\hat{A}$-genus vanishes.

\medskip
The Atiyah-Bott-Shapiro map 
$M{\rm Spin}\to K{\rm O}$
is constructed using the representations of the Spin groups, and relies on knowing that for a space $X$, elements 
of $K{\rm O}^0(X)$ are represented by vector bundles over $X$. It gives a $K{\rm O}$-theory Thom isomorphism 
for Spin-vector bundles, and is  a topological expression for the index of the Dirac operator.
We hope that similar questions 
can be explored in the semi-Riemannian setting.

  \medskip
 This paper is organized as follows.
We start with a more general setting of the problem that we hope 
explain some of the homotopy theoretical constructions encountered 
here as well as in previous literature. In Sec. \ref{Sec Postnikov} 
and \ref{Sec Whitehead}, we describe the Postnikov tower and the Whitehead tower
of a space in a manner that is appropriate for applications. 
Then in Sec. 
\ref{Sec Variant}
we provide a variant point of view on the Whitehead tower in way of clarification. 
Since lifts of indefinite Lie groups will be determined by their maximal compact subgroups, 
which are products, we discuss in Sec. \ref{Sec Computational}
useful conditions on behavior of cohomology of products. The discussion is 
needed since we work with integral cohomology. 
This is applied in Sec. \ref{Sec Special} to classifying spaces, where we 
identify the obstructions and where the fibrations become fiber bundles. 
We start with the applications in Sec. 
\ref{Sec app}.
First, in order to make sure we are on firm ground, we discuss the indefinite 
orthogonal groups in Sec.
\ref{Sec SO}, highlighting their unstable homotopy groups. A subtle issue in 
the non-vanishing of the fundamental group of the corresponding Spin groups is 
addressed in Sec. \ref{Sec Spin}. 
This is used in Sec. \ref{Sec String} to define the indefinite String groups, where
we identify the obstructions explicitly by studying the generators of the classifying space of 
$B{\rm Spin}(n)$ in the unstable case. 
We then show in Sec. \ref{Sec U Sp} how the definitions and constructions extend to the
case of ${\rm U}(p, q)$ and ${\rm Sp}(p, q)$, where subtle issues with stability are absent. 
Finally, in Sec. \ref{Sec Twisted} we describe relations to twisted structures, and 
end with highlighting future work that we hope to do. 

\section{The Postnikov tower, Whitehead tower, and variants}

In this section we will provide a careful treatment of the towers arising in co-killing homotopy groups of Lie groups.
The idea is that Postnikov towers arise when killing homotopy groups above a certain degree, while
Whitehead towers arise when killing homotopy groups below a certain degree-- hence the term co-killing.
We will also provide a variant tower of higher connected covers, which we demonstrate is equivalent to the 
latter. 
We believe that such a treatment, while certainly known to experts, seems to be missing from existing 
literature to the best of our knowledge. This technical treatment, we hope, will be for the benefit of the 
reader and will make the paper self-contained. Readers not interested in these details might wish to skip
this section.

\subsection{The Postnikov tower} 
\label{Sec Postnikov} 

We start by recalling the Postnikov tower (see \cite{Hu} \cite{Hat}). 
We will take this as our starting point to connect to the Whitehead tower.

\begin{theorem} (see \cite{Hu} \cite{Hat}) Let $X$ be a simple path-connected space with a 
map $\alpha:X\to X_1$, where $X_1$ is the Eilenberg-MacLane 
space $K(\pi_1(X),1)$, that induces an isomorphism on fundamental groups 
$\pi_1(X)\to \pi_1(X_1)$. 
Then there are spaces $X_n$ with maps $\alpha_n:X\to X_n$ 
which induce isomorphisms on homotopy groups
$\pi_k(X)\to \pi_k(X_n)$ for $k\le n$ and $\pi_k(X_n)=0$ 
for $k>n$ with fibrations $p_{n+1}:X_{n+1} \to X_n$ such that 
$\alpha_{n}=p_{n+1}\circ \alpha_{n+1}$. 
\end{theorem}
Assuming that such $\alpha_n:X\to X_n$ is given, the space $X_{n+1}$ 
will be defined as the homotopy fiber
 of a certain map $k^{n+2}:X_n \to K(\pi_{n+1}(X),n+2)$ that is equivalent to the corresponding cohomology class 
 in $H^{n+2}(X_n,\pi_{n+1}(X))$ such that the induced class $\alpha_n^\ast k^{n+2}\in H^{n+2}(X,\pi_{n+1}(X))$ is trivial.
By the universality of the homotopy fiber $X_{n+1}$, one then obtain the desired map 
$\alpha_{n+1}:X\to X_{n+1}$.
These maps are assembled in the following diagram called the \emph{Postnikov tower} of $X$:
\(\xymatrix{
&&\vdots \ar[d] &\\
&& X_n \ar[r]^-{k^{n+2}} \ar[d]^-{p_n}& K(\pi_{n+1}(X),n+2)\\
&& \vdots \ar[d]^-{p_3} &\\
&& X_2 \ar[r]^-{k^4} \ar[d]^-{p_2} & K(\pi_3(X),4)\\
X \ar[rr]^-{\alpha_1} \ar[urr]^-{\alpha_2} \ar[uuurr]^-{\alpha_n} 
&& X_1 \ar[r]^-{k^3} & K(\pi_2(X),3)\;.
}\)
One could see how to choose such a map $k^{n+2}:X_n \to K(\pi_{n+1}(X),n+2)$ and 
how to obtain $X_{n+1}$ as the homotopy fiber explicitly as follows. Let 
$j:C(\alpha_n)\to K(\pi_{n+1}(X), n+2)$ be the inductive attachment of cells to the cofiber 
$C(\alpha_n)$ to match the homotopy group of the space $K(\pi_{n+1}(X), n+2)$ from the identification 
$\pi_{n+2}(C(\alpha_n))\cong \pi_{n+1}(X)$ and the fact that $C(\alpha_n)$ is $(n+1)$-connected. On the other 
hand, we have the inclusion $X_n \hookrightarrow C(\alpha_n)$ and we take the composite 
$k^{n+2}:X_n\to K(\pi_{n+1}(X),n+2)$.
From the inclusion of the cone $C(X)$ in the cofiber $C(\alpha_n)$, $\chi_x(t):=j(x,1-t)$ for $(x,1-t)\in C(X)$ is in
the path space 
 $PK(\pi_{n+1}(X),n+2)$. This then makes the outer rectangle part of the following pullback diagram 
\(
\xymatrix{
X \ar@/^2em/[drrr]^{x\mapsto \chi_x} \ar@/_2em/[ddr]_{\alpha_n} \ar@{-->}[dr]^{\alpha_{n+1}}&&&
\\
&X_{n+1}  \ar[d]_{p_{n+1}} \ar[rr]&& PK(\pi_{n+1}(X),n+2)\ar[d]^{\mathrm{ev}_1} \\
&X_n \ar[rr]^-{k^{n+2}} && K(\pi_{n+1}(X),n+2)
}
\)
commutative, so that we have the homotopy fiber $X_{n+1}$. Here ${\rm ev}_1$ denotes 
evaluation at the point 1 of the interval in the path space. Note that $PK(\pi_{n+1}(X),n+2)$ is 
a homotopy equivalent to a point space so the commutativity of the diagram implies that the 
cohomology class $\alpha_n^\ast k^{n+2}\in H^{n+2}(X;\pi_{n+1}(X))$ is trivial.
Therefore, we obtain the map $\alpha_{n+1}$ which satisfies $\alpha_n=p_{n+1}\circ \alpha_{n+1}$
 and induces an isomorphism $\pi_k(X)\to \pi_k(X_{n+1})$ for $k\le n+1$ and makes 
 $X_{n+1}$ have trivial $k^\mathrm{th}$ homotopy group for $k>n+2$. By inductive application 
 of this process, one obtains the \emph{Postnikov tower} of $X$.


\subsection{The Whitehead tower}
\label{Sec Whitehead}

Next we consider the Whitehead tower, which is in a sense a dual to the Postnikov tower \cite{Wh} \cite{Whi}. 
Note that the Postnikov tower approximates the homotopy groups of $X$ ``from the bottom'' in the sense that
 it admits an isomorphism of lower homotopy groups with higher homotopy groups being killed. One then may
  seek  the dual process of estimating the homotopy groups of $X$ ``from the top'', in the sense that lower 
  homotopy groups are killed while  higher homotopy groups are isomorphic to those of $X$.

\begin{theorem} (See \cite{Whi}) Let $X$ be a (path) connected space with a Postnikov tower.
Then there are (path) connected spaces ${X}\langle n\rangle$ such that $\pi_k({X}\langle n\rangle)=0$ for $k\le n$ 
and maps $\widehat{\alpha}_n: {X}\langle n\rangle\to X$ that induce isomorphisms 
$\pi_k({X}\langle n\rangle)\to \pi_k(X)$ for $k> n$.  Moreover, there is a fibration 
$\widehat{p}_{n+1}: {X}\langle {n+1}\rangle\to {X}\langle n\rangle$ for each $n$ such that 
$\widehat{\alpha}_{n+1}=\widehat{\alpha}_n\circ\widehat{p}_{n+1}$   with a fiber the 
based loop space $\Omega X\langle n\rangle$.
\end{theorem}

From the Postnikov tower of $X$, we have spaces 
$X_n$ and maps $\alpha_n: X \to X_n$. By taking the homotopy fiber 
${X}\langle {n}\rangle$ of $\alpha_n$, we obtain a fibration
$\widehat{\alpha}_n: {X}\langle n\rangle \to X$ that induces isomorphisms 
$\pi_k({X}\langle {n}\rangle)\to \pi_k(X)$ for $k>n$ and makes 
$\pi_k({X}\langle n\rangle)=0$ for $k \le n$.
Since ${X}\langle {n+1}\rangle \to X$ factors through ${X}\langle n\rangle$ naturally as in the diagram
\(
\xymatrix{
{X}\langle {n+1}\rangle\ar@/^/[r] \ar@/_1em/[ddr]_{\widehat{\alpha}_{n+1}} 
\ar@{-->}[dr]^{\widehat{p}_{n+1}}&PX\langle {n+1}\rangle\ar@/^/[dr]^{P(p_{n+1})}&\\
&{X}\langle n\rangle \ar[r] \ar[d]_{\widehat{\alpha}_n} & PX_n\ar[d]\\
&X\ar[r]^-{\alpha_n} & X_n\;,
}
\)
we obtain a map $\widehat{p}_{n+1}:{X}\langle {n+1}\rangle\to {X}\langle n\rangle$, which can be made into
 a fibration, up to homotopy equivalence, such that $\widehat{\alpha}_{n+1}=\widehat{\alpha}_n\circ \widehat{p}_{n+1}$.
The induced long exact sequence of the fibration $\widehat{p}_{n+1}$ shows that the fiber is homotopy equivalent to
 the Eilenberg-MacLane space $K(\pi_{n+1}(X), n)$.
The tower of fibrations thus obtained is a dual version of the Postnikov tower called the \emph{Whitehead tower} \cite{Wh} \cite{Whi}:
\begin{equation}\label{201411161345}
\xymatrix@C=5em{
&\vdots \ar[d] &&&\\
K(\pi_{n}(X),{n-1}) \ar[r] & {X}\langle n\rangle \ar[ddddrr]^{\widehat{\alpha}_n} \ar[d]_-{\widehat{p}_{n}} && \\
& \vdots \ar[d] && \\
K(\pi_3(X),2) \ar[r] & {X}\langle 3\rangle \ar[ddrr]^{\widehat{\alpha}_3} \ar[d]_{\widehat{p}_{3}} &&\\
K(\pi_2(X),1) \ar[r] & {X}\langle 2\rangle \ar[drr]^{\widehat{\alpha}_2} \ar[d]_{\widehat{p}_{2}} && \\
& {X}\langle 1\rangle \ar[rr]^{\widehat{\alpha}_1} && X.
}\end{equation}

\subsection{A variant view}
\label{Sec Variant}

Each fibration $\widehat{p}_{n+1}: {X}\langle {n+1}\rangle \to {X}\langle n\rangle$ in the Whitehead tower 
can be regarded as the $(n+1)$-connected covering of ${X}\langle n\rangle$ and we may ask the 
``lifting" question, i.e. under what circumstance can a map $M\to {X}\langle n\rangle$
be lifted to $M\to {X}\langle {n+1}\rangle$ as in the commutative diagram
\(
\xymatrix{
& {X}\langle {n+1}\rangle \ar[d]\\
M \ar[r] \ar@{-->}[ur] & {X}\langle n\rangle\;.
}\)
To answer this question we need the following important property 
of the sequence of spaces appearing in the Whitehead tower.

\begin{proposition} 
Let $X$ be a connected space. Then there are spaces $X\langle n\rangle$, where $X\langle 1\rangle:=X$, 
with fibrations $\hat{p}_{n+1}: X\langle n+1\rangle\to X\langle n\rangle$ with fiber $K(\pi_n(X),n-1)$ such that 
$$
\pi_k(X\langle n\rangle)
\cong 
\left\{
\begin{tabular}{lc} 
$0$ & {\rm for} $k<n$ \\
$\pi_k(X)$ & {\rm for} $k\geq n$.  
\end{tabular}
\right.
$$
\end{proposition}

\begin{remark}
These spaces $X\langle n+1\rangle$ are constructed as the homotopy fibers of a cohomology class
 $\lambda_n\in H^n(X\langle n\rangle;\pi_n(X))$ 
that induces an isomorphism $\pi_n(X\langle n\rangle)\overset{\cong}{\to} \pi_n(X)$, and a map $f:M\to X\langle n\rangle$
 is lifted to $\widetilde{f}:M\to X\langle n+1\rangle$ if the induced cohomology class $f^\ast \lambda_n\in H^n(M;\pi_n(X))$ 
 is trivial.
 \end{remark}

\noindent Thus we have the following diagram that one might call the \emph{tower of higher connected covers}
\(\xymatrix{
&&\vdots \ar[d] &\\
&& X\langle n\rangle \ar[r]^-{\lambda_{n}} \ar[d]^-{q_n}& K(\pi_{n}(X),n)\\
&& \vdots \ar[d]^-{q_3} &\\
&& X\langle 2\rangle \ar[r]^-{\lambda_2} \ar[d]^-{\lambda_2} & K(\pi_2(X),2)\\
M \ar[rr] \ar[urr] \ar@{-->}[uuurr] && X\langle 1\rangle \ar[r]^-{\lambda_1} & K(\pi_1(X),1)\;.
}\)
The construction is an inductive application of the process in the following lemma.
\begin{lemma} [Killing $n^\textrm{th}$ homotopy group] Let $X$ be a simply connected space and 
$Y$ be an $(n-1)$-connected space  whose cohomology class $\lambda_n\in H^n(Y;\pi_n(X))$ 
induces an isomorphism $\pi_n(Y)\to \pi_n(X)$. Then the homotopy fiber $\widehat{Y}$ of $\lambda_n$ satisfies
$$
\pi_k(\widehat{Y})
\cong 
\left\{
\begin{tabular}{lc} 
$0$ & {\rm for} $k=n$ \\
$\pi_k(Y)$ & {\rm for} $k\neq n$,
\end{tabular}
\right.
$$
with a fibration $q:\widehat{Y}\to Y$. A map $f:M\to Y$ is lifted 
to $\widehat{f}:M\to \widehat{Y}$ with respect to the fibration $q$ when the induced cohomology class $f^\ast \lambda_n\in H^n(M;\pi_n(X))$ is trivial.
\end{lemma}
\proof
We would like to consider the lifting 
\(
\xymatrix{
& \widehat{Y} \ar[d]^q & \\
M \ar@{-->}[ur]^{\widehat{f}} \ar[r]^f & Y \ar[r]^-{\lambda_n}& K(\pi_n(X), n)\;. 
}
\)
The result then follows
from considering the following pullback diagram  
\begin{equation}\label{Pull diag1}\xymatrix{
M\ar@{-->}[dr]^{\widehat{f}} \ar@/_1em/[ddr]_{f} \ar@/^1em/[drr]^{} && \\
& \widehat{Y} \ar[d]_{q}\ar[r]  & PK(\pi_n(X),n))\ar[d]^{\mathrm{ev}_1}\\
& Y \ar[r]^-{\lambda_{n}}& K(\pi_n(X),n)
}
\end{equation}
 and its long exact sequence of homotopy groups and the fact that the path space 
 $PK(\pi_n(X),n)$ is contractible. Here, the map ${\rm ev}_1$ is the evaluation 
 map at the point 1 in the path. 
\endofproof

\begin{remark} {\bf (i)}  
One way of figuring out that  a given fibration $K(\pi_n(X),n-1)\to X\langle n+1\rangle \to X\langle n\rangle$ fits in 
the tower of higher connected covers is to check that a map $f:M\to X\langle n\rangle$ can be lifted up to a map 
$\widehat{f}:M\to X\langle n+1\rangle$ if and only if the cohomology class $f^\ast \lambda_n\in H^n(M;\pi_n(X))$ 
induced by $\lambda_n$ which induced the isomorphism of $n^\textrm{th}$-homotopy group vanishes.

\vspace{1mm}
\noindent {\bf (ii)} It is convenient to rework the definition in a way making explicit the obstruction problem 
giving the lift from $X\langle n-1 \rangle$ to $X\langle n \rangle$. Indeed,  the two 
constructions, namely the Whitehead tower and the tower of higher connected covers, 
are equivalent by the 
homotopy commutative diagram:
\(
\xymatrix{
X\langle n \rangle \ar[d] \ar[rr] && \ast \ar[d] && \\
X \langle n-1 \rangle \ar[d] \ar[rr] && K(\pi_n(X), n) \ar[d] \ar[rr] && \ast \ar[d]\\
X \ar[rr] && X\langle n\rangle \ar[rr] && X\langle {n-1}\rangle\;.
}
\)
\end{remark}

\subsection{Computational aspects} 
\label{Sec Computational}

We now turn to useful computational aspects of the above constructions. This will also be useful when considering 
indefinite Lie groups in Sec. \ref{Sec app}. 
We would like to determine which classes in $H^n(Y;\pi_n(X))$, realized as 
homotopy classes of maps $Y \to K(\pi_n(X), n)$, induce  isomorphisms $\pi_n(Y)\cong \pi_n(X)$.
We do so by relating the summand ${\rm Hom}(H_n(Y, \Z), \pi_n(X))$ in 
 $H^n(Y;\pi_n(X))$ via the Hurewicz map $\pi_n(Y) \to H_n(Y, \Z)$. In general, one 
 needs strong assumptions on $Y$ in order for this to be viable, as the Hurewicz map could be 
 zero. However, under the assumption that $Y$ is $(n-1)$-connected, we have
 an isomorphism $H^n(Y, \pi_n(X)) \cong {\rm Hom}(\pi_n(Y), \pi_n(X))$. 


\medskip
Suppose that $n{\rm th}$ homotopy group of $Y$ is of the form $A\times B$
for some abelian groups $A$ and $B$ such that each of $H^n(Y;A)$ and $H^n(Y;B)$ is freely generated by 
a single generator $\alpha$ and $\beta$, respectively. 
We will denote by $(\alpha, 0)$ and $(0, \beta)$ the classes in $H^n(Y, A \times B)$ obtained from the 
corresponding classes in $H^n(Y; A)$ and $H^n(Y; B)$ via the canonical homomorphism 
$A \to A \times B$ and $B \to A \times B$, respectively. 
Then $\alpha$ and $\beta$ induce group homomorphism 
$\pi_n(X)\to A$ and $\pi_n(X)\to B$ so that neither $(\alpha,0)$ nor $(0,\beta)$ in $H^n(Y;A\times B)$ can 
induce an isomorphism $\pi_n(Y)\buildrel{\cong}\over{\to} A\times B$. 
However, the class in $H^n(Y;A)\times H^n(Y;B)$ inducing the isomorphism does exist and is of the form 
$(\tilde{\alpha},\tilde{\beta})$ where $\tilde{\alpha}=a\alpha\in H^n(Y;A)$ and $\tilde{\beta}= b \beta\in H^n(Y;B)$ for some nonzero integers $a$ and $b$.
At any rate, taking $Y$ is $(n-1)$-connected, we have an isomorphism
$$
{\rm Hom}_\Z(\pi_n(Y), A) \times {\rm Hom}_\Z(\pi_n(Y), A) \longrightarrow {\rm Hom}_\Z(\pi_n(Y), A\times B)\;.
$$
On the other hand, still assuming $\pi_n(X) \cong \pi_n(Y)$, 
if $Y$ is homotopy equivalent to the product space $Y'\times Y''$ and
\begin{equation} \label{201410271934}
H^n(Y;\pi_n(X))\cong H^n(Y';\pi_n(X))\times H^n(Y'';\pi_n(X))
\end{equation}
then we have the decomposition 
\(
H^n(Y;\pi_n(X))\cong H^n(Y';\pi_n(Y'))\times H^n(Y';\pi_n(Y''))\times H^n(Y'';\pi_n(Y'))\times H^n(Y'';\pi_n(Y''))\;.
\)
This is immediately the case for our spaces, since they are assumed to be $(n-1)$-connected. 
%
%

\medskip
Assuming further that both abelian groups $H^n(Y';\pi_n(Y'))$ and $H^n(Y'';\pi_n(Y''))$ are 
each cyclic with a single class generator  $\alpha'$ and $\alpha''$, respectively, 
the cohomology class $(\alpha',\alpha'')$ induces the desired isomorphism 
$\pi_n(Y'\times Y'')\buildrel{\cong}\over{\to} \pi_n(Y')\times \pi_n(Y'')$.
Obviously, the cohomology class $(\alpha',\alpha'')$ is represented by
 $\alpha'\times\alpha'':Y'\times Y''\to K(\pi_n(Y'),n)\times K(\pi_n(Y''),n)$.
Thus, we have obtained:

\begin{proposition}
 \label{prop BG} For a path-connected and $(n-1)$-connected space $Y$ 
the homotopy fiber $\widehat{Y}$ of $\lambda_n:Y\to K(\pi_n(Y),n)$ has homotopy 
groups isomorphic to those of $Y$ except $\pi_n(\widehat{Y})=0$, according to the following cases:
\begin{enumerate}
\item[{\bf (i)}] $\lambda_n = \alpha$ if $\alpha$ is the only generator of $H^n(Y,\pi_n(X))$, i.e.,
we have a homotopy pullback diagram 
$$\xymatrix{
\widehat{Y}\ar[d] \ar[r] & {\rm pt}\ar[d]\\
Y \ar[r]^-{\alpha}& K(\pi_n(X),n)\;.
}
$$
\item[{\bf (ii)}] $\lambda_n = (\widehat{\alpha},\widehat{\beta})$ if $\pi_n(Y)\cong A\times B$ and 
 $H^n(Y;A)$ and $H^n(Y;B)$ are free with generators $\alpha$ and $\beta$, respectively, 
 where $\widehat{\alpha}=a\alpha$ and 
$\widehat{\beta}=b\beta$ for some nonzero integers $a$ and $b$, respectively, i.e. 
$$\xymatrix{
\widehat{Y}\ar[d] \ar[rr] && {\rm pt}\ar[d]\\
Y \ar[rr]^-{(\widehat{\alpha},\widehat{\beta})}&& K(A\times B,n)\;.
}$$

\item[{\bf (iii)}] $\lambda_n = \alpha'\times \alpha''$ which is equivalent to the map 
$Y'\times Y''\to K(\pi_n(Y'),n)\times K(\pi_n(Y''),n)$ if $Y\simeq Y'\times Y''$ where $\alpha',\alpha''$ 
are the only generators of $H^n(Y',\pi_n(Y'))$ and $H^n(Y'',\pi_n(Y''))$ respectively, 
and we again have the homotopy pullback diagram  
$$\xymatrix{
\widehat{Y}\ar[d] \ar[rr] && {\rm pt}\ar[d]\\
Y'\times Y'' \ar[rr]^-{\alpha'\times\alpha''}&& K(\pi_n(Y')\times\pi_n(Y''),n)\;.
}$$
\end{enumerate}
\end{proposition}

\subsection{Special case: higher connected covers of classifying spaces and obstructions}
\label{Sec Special}

We will mainly be interested in our spaces being topological groups. Let $G$ be an $(n-1)$-connected
 topological group.
 \footnote{Previously we denoted this by $G\langle n\rangle$ but for ease of notation we drop 
 the extra decoration here.} 
Then $BG$ is a classifying space of $G$ such that $\pi_{i}(BG)=0$ for $i \leq n$.
By taking the homotopy fiber of the canonical map $BG\to K(\pi_{n+1}(BG),n+1)$, we obtain a space
$\widehat{BG}$ for which $\pi_{n+1}(\widehat{BG})$ is trivial. Then its loop space $\Omega\widehat{BG}$ 
satisfies $\pi_{n}(\Omega\widehat{BG})=0$. 
Setting $\widehat{G}:=\Omega \widehat{BG}$, we get a topological space which is $G$ with $\pi_n$ killed. 
The immediate 
consequence of this observation is the following:
\begin{lemma} $\widehat{BG}$ is weakly homotopy equivalent to $B(\widehat{G})$.
\end{lemma}

Note that the  statement of the lemma can be promoted to a homotopy equivalence provided $B(\widehat{G})$ is taken
to be the usual classifying space of the (group-like) $A_\infty$-space $\widehat{G}$.

\medskip
We now consider classification of the corresponding bundles. From the general discussion in previous sections
we have the following results. 

\begin{proposition} Suppose that there is a $G$-principal bundle over $M$ determined by a classifying map 
$f:M\to BG$. Then there exists a map $\hat{f}:M\to B\widehat{G}$ providing a $\widehat{G}$-principal bundle over 
the same base $M$ compatible with the bundle $f$ with respect to the map $B\widehat{G}\to BG$
if $f^\ast\lambda_{n}\in H^{n+1}(M,\pi_{n}(G))$ is trivial for some $\lambda_n\in H^{n+1}(BG;\pi_{n}(G))$
 that induces an isomorphism $\pi_n(BG)\to \pi_{n-1}(G)$.
\label{Prop Gprinc}
\end{proposition}

\begin{remark}
{\bf (i)} The condition in Proposition \eqref{Prop Gprinc}
 is equivalent to saying that $f^\ast \lambda_n$ factors through a point space up to homotopy:
\(\xymatrix{
M\ar@/_1.5em/[ddr]_{f} \ar@/^1.5em/[drr] \ar@{-->}[dr]^{\widehat{f}} &&\\
& B\widehat{G}\ar[r]\ar[d]& {\rm pt}\ar[d]\\
& BG \ar[r]^-{\lambda_n}& K(\pi_{n}(G),n+1)\;.
}\)

\vspace{1mm}
\noindent {\bf (ii)} We say that the \emph{obstruction} $f^\ast \lambda_{n}$ \emph{vanishes} or \emph{trivialized} so
 that the bundle $f$ is lifted to $\hat{f}$.
This lifting of $f$ to $\hat{f}$ is also called the \emph{trivialization} of the class $\lambda_n$.
\end{remark}

\medskip
Indeed, the pullback diagram eq. \eqref{Pull diag1} induces another pullback diagram classifying 
$\widehat{G}$-principal bundles over $X$. Denoting the set of  isomorphism classes of $G$-principal 
bundles over $X$ by ${\rm Bun}_G(X)$, we have a pullback diagram
\(\xymatrix{
{\rm Bun}_{\widehat{G}}(X) \ar[r]  \ar[d]& {\rm pt}\ar[d] \\
{\rm Bun}_{G}(X) \ar[r] & H^{n}(X;\pi_{n-1}(G))\;.
}\)
Inductively, we get a tower of connected covers which is the right side of the diagram:
\begin{equation}\label{201411041232} \xymatrix@R=1.5em{
&&\vdots\ar[d]&\vdots\\
&& BG\langle n\rangle\ar[d] \ar[r]^-{\lambda_{n}} & K(\pi_{n-1}(G),n)\\
& &\vdots\ar[d]^-{q_2} &\vdots \\
&& BG\langle 2\rangle\ar[d]^-{q_1}  \ar[r]^-{\lambda_2} & K(\pi_1(G),2)\\
&& BG\langle 1\rangle \ar[r]^-{\lambda_1}\ar[d]^{q_0}& K(\pi_0(G),1)\\
X\ar[rr]
\ar[urr] 
\ar[uurr]
\ar[uuuurr]
&& BG \ar[r]^-{\lambda_0}& K(\pi_0(BG),0)\;.
}\end{equation}

\medskip
More particularly, we will be mainly interested in the case when $G\langle n\rangle$ is an $(n-1)$-connected 
topological group, and also in the classifying space $BG$ for some topological group $G$, i.e., $X=BG$, and 
its $(n-1)$-connected cover $BG\langle n\rangle$.  
We would like to find an $(n-2)$-connected topological group  $G\langle n-1\rangle$ forming part of 
an $(n-2)$-connected cover $G\langle n-1\rangle \to G$ such that $B(G\langle n-1\rangle)$ is 
homotopy equivalent to  $(BG )\langle n\rangle$. 

\medskip
For relatively low 
\footnote{The arguments work rationally for any $k$ \cite{FSSt}\cite{SW}. However in the integration
from algebras to groups in the $L_\infty$ setting for the orthogonal case, one encounters difficulty for the
 $\Z/2$ groups in the homotopy groups of $O$. So our discussions work fully for String and Fivebrane,
 which is more than what we need for this paper. For the complex/unitary case, the homotopy groups 
 are torsion-free and so no such difficulties arise.} 
$n$ the connected covers ${\rm O}(k)\langle n\rangle$ are defined as the based loop 
spaces of the corresponding classifying spaces in \cite{SSS3}\cite{FSSt}\cite{9brane}. It follows from 
the results of Kan and Milnor that every based loop space has the homotopy type of a topological group. 
In the homotopy category of connected CW complexes, there is an equivalence between loop spaces, 
topological groups, and associative H-spaces (see \cite[Ch. 4]{Ka}). 
Rationally, there is essentially a unique multiplication on the connected covers of the 
orthogonal group \cite[Prop. 3]{SW}.

\medskip
We have  the homotopy pullback diagram 
\(
\xymatrix{
\widehat{BG}\langle n \rangle \ar[r] \ar[d] & {\rm pt}\ar[d]\\
BG \langle n \rangle\ar[r] & K(\pi_{n}(G),n+1)
}
\label{Hom pull diag}
\)
and the fibration 
\(
K(\pi_n(G),n)\longrightarrow \widehat{BG}\langle n \rangle \longrightarrow BG\langle n \rangle\;.
\label{fib1}
\)
Having a homotopy pullback diagram such as \eqref{Hom pull diag}
is equivalent to saying that $\widehat{BG}\langle n \rangle$ is the homotopy fiber of 
$BG\langle n \rangle \to K(\pi_{n}(G),n+1)$, hence equivalent to saying that 
$\widehat{BG}\langle n \rangle \to BG\langle n \rangle \to K(\pi_{n}(G),n+1)$ is a homotopy fiber sequence.

\medskip
Next, setting $\widehat{G}\langle n-1 \rangle=\Omega(\widehat{BG}\langle n \rangle)$ 
we have $B(\widehat{G}\langle n-1 \rangle)\simeq \widehat{BG}\langle n \rangle$,
 and this $\widehat{G}\langle n-1 \rangle$ fits into the diagram
\(\xymatrix{
\widehat{G}\langle n-1 \rangle \ar[r] \ar[d] & {\rm pt}\ar[d]\\
G\langle n-1 \rangle \ar[r] & K(\pi_n(G), n)\;,
}\)
which induces a fibration with fiber $K(\pi_n(G),n-1)$
\begin{equation}
\label{KGG}
K(\pi_{n}(G),n-1)\longrightarrow \widehat{G}\langle n-1 \rangle\longrightarrow G\langle n-1 \rangle\;.
\end{equation}
Since $K(\pi_n(G),n-1)$ has a group structure, then this induces \eqref{fib1} as a fiber bundle.
In fact, this will a {\it principal fiber bundle}, the point being that the homotopy
fiber of a map to a connected space $X$ is actually an $\Omega X$-principal
bundle with the structure group an actual topological group
representing the homotopy type of the loop space $\Omega X$.

\medskip
For example, for the case $n=1$, we have $\widehat{G}={\rm Spin} (m)$ defined in terms of Clifford algebra 
to be the double cover of $G={\rm SO}(m)$, $m\geq 3$.


\medskip
Let $G$ be an $n$-connected topological group with $n \geq 2$. If $\pi_{n+1}(G)=A$, then 
there is a model for the Eilenberg-MacLane space $K(A, n)$ with the structure of a 
topological abelian group which forms part of an extension of topological abelian groups
$$
1 \longrightarrow K(A, n) \longrightarrow \widehat{G} \xrightarrow{\;\;\rho\;\;} G \longrightarrow 1\;.
$$
Moreover, $\widehat{\rho}: \widehat{G} \to G$ is a principal $K(A, n)$-bundle.
This can be proved similarly to the $n=2$ case in \cite{St}, by choosing a model for $K(A, n)$ as an 
abelian topological group and letting $P \to G$ denote the principal $K(A, n)$-bundle
classified by the fundamental class of $H^{n+1}(G, A)$. Then define $\widehat{G}$
as in \cite{St} so that we have a short exact sequence of topological groups
$$
1 \longrightarrow {\rm Gauge}(P) \longrightarrow \widehat{G} \longrightarrow G \longrightarrow 1\;.
$$
where ${\rm Gauge}(P)$ is the gauge group of the bundle $P$. Similarly to \cite{St} one
can show that the canonical evaluation homomorphism ${\rm Gauge}(P) \to K(A, n)$ is a homotopy
equivalence. In summary, we have:

\begin{proposition}
The fibrations in the Whitehead tower (hence also the tower of connected covers) of a Lie group
are principal fiber bundles. 
\end{proposition}

Note that a detailed treatment of the Whitehead tower in the rational case is given 
in \cite{SW}, where the group structures are also identified. 



\section{Applications} 
\label{Sec app}

\subsection{Special orthogonal groups ${\rm SO}(n)$ and ${\rm SO}(p, q)$}
\label{Sec SO}

We now concentrate on the orthogonal group and its connected covers. We start with the 
definite signature and then work our way to  indefinite signatures. Starting with the former, 
we have the statement that 
identifies when the shift in rank is inconsequential 
\begin{equation} 
\pi_i({\rm O}(n))\cong \pi_i({\rm O}(n+1)) \quad \text{ for } 0< i\le n-2\;.
\end{equation}
Note that according to Kervaire \cite{Ker}, this cannot be improved in general as
$\pi_6({\rm SO}(7))\cong 0$, while  $\pi_7({\rm SO}(7))\cong \Z$.
The following isomorphism shows that the special orthogonal group ${\rm SO}(n)$
has the same homotopy groups as the orthogonal group ${\rm O}(n)$ in 
positive degrees
\(
\pi_i({\rm SO}(n))=\begin{cases}
0 & \text{ for }i=0\\
\pi_i({\rm O}(n))&\text{ for }i\geq 1.
\end{cases}
\)
We also note the following low degree identifications, which are often useful in calculations and applications:
${\rm O}(1)\cong S^0$, ${\rm SO}(2)\cong S^1$, ${\rm SO}(3)\cong \R P^3$, while 
${\rm SO}(4)$ is the double cover of ${\rm SO}(3) \times {\rm SO}(3)$, i.e.,  
${\rm SO}(4)\cong (S^3\times S^3)/\Z_2$. Note that ${\rm Spin}(4) \cong S^3 \times S^3$.

\medskip
Collecting all the above observations and results
 gives the following table of {\it unstable} homotopy groups of lower dimensional orthogonal groups
 (see e.g. \cite{Ker}): 

\begin{center}\begin{tabular}{ |c || c c c c c c c c c|}
\hline
&${\rm O}(1)$& ${\rm O}(2)$ & ${\rm O}(3)$ & ${\rm O}(4)$ & ${\rm O}(5)$ & ${\rm O}(6)$ &${\rm O}(7)$ & ${\rm O}(8)$ & ${\rm O}(9)$\\
\hline
$\pi_0$ & $\Z/2$ & $\boxed{\Z/2}$ & $\Z/2$ & $\Z/2$ & $\Z/2$ & $\Z/2$ & $\Z/2$ & $\Z/2$ & $\Z/2$\\
$\pi_1$ & $0$ & $\Z$ & $\boxed{\Z/2}$ & & & &&& \\
$\pi_2$ & $0$ & $0$& $0$ & $\boxed{0}$&  & &&& \\
$\pi_3$& $0$ & $0$ & $\Z$ & $\Z\times\Z$ & $\boxed{\Z}$ &&&&\\
$\pi_4$& $0$ & $0$ & $\Z/2$ & $\Z/2\times\Z/2$& $\Z/2$ & $\boxed{0}$&&&\\
$\pi_5$& $0$ & $0$ & $\Z/2$ & $\Z/2\times\Z/2$& $\Z/2$ & $\Z$&$\boxed{0}$&&\\
$\pi_6$& $0$ & $0$ & $\Z/12$ & $\Z/12\times\Z/12$& $0$ & $0$&$0$ &$\boxed{0}$&\\
$\pi_7$& $0$ & $0$ & $\Z/2$ & $\Z/2\times\Z/2$& $\Z$ & $\Z$&$\Z$&$\Z\times\Z$& $\boxed{\Z}$\\
\hline
 \end{tabular}
 \end{center}
 
\vspace{2mm} 
\noindent where the boxed entries indicate that the corresponding homotopy groups are being stabilized.

\medskip
We now consider the corresponding classifying spaces. 
We know that the first cohomology $H^1(B{\rm O}(n),\Z/2)\cong \Z/2$ is generated by the first 
Stiefel-Whitney class $w_1$,  so that we may pullback $w_1:B{\rm O}(n)\to K(\Z/2,1)$ to obtain 
a $0$-connected cover, say $G$, of ${\rm O}(n)$:
\(
\xymatrix{
& BG \ar[r] \ar[d] & {\rm pt} \ar[d]\\
X\ar[r]^-f \ar[ur]^{\widehat{f}}& B{\rm O}(n)\ar[r]^-{w_1} & K(\Z/2,1)\;.
}
\)
We also know that the special orthogonal group ${\rm SO}(n)$ has the same homotopy groups 
as $G$ so we may regard the connected identity component ${\rm SO}(n)$ of ${\rm O}(n)$ as 
the $0$-connected cover $G$. The pattern continues to the Spin group and beyond, as in \cite{SSS2}. 

\medskip
Next, consider our main object which is the indefinite orthogonal group ${\rm O}(p, q)$. This has 
a maximal compact subgroup ${\rm O}(p)\times {\rm O}(q)$. 
The inclusion ${\rm SO}(p) \times {\rm SO}(q) \hookrightarrow {\rm SO}(p, q)$ 
is a homotopy equivalence by the Cartan decomposition of noncompact Lie groups,
i.e. the homeomorphism from ${\rm SO}(p, q) \to {\rm SO}(p) \times {\rm SO}(q) \times \R^{pq}$.

\medskip
With this observation, we 
can reduce the problem of connected covers in this indefinite signature setting to essentially two copies of the problem in the 
definite signature case. From the results in the previous section, we may pullback 
$w_1\times w_1:B{\rm O}(p)\times B{\rm O}(q)\to K(\Z/2,1)\times K(\Z/2,1)$ to obtain
 the 0-connected cover $G$:
\(
\xymatrix{
&BG(p,q) \ar[rr] \ar[d] &&{\rm pt}\ar[d]\\
X\ar[r]_-{f}\ar[ur]^{\widehat{f}}&B{\rm O}(p,q)\ar[rr]^-{w_1\times w_1} && K(\Z/2\times \Z/2,1)\;.
}
\)
We have $G(p,q)\simeq{\rm SO}(p)\times{\rm SO}(q)$ and this is homotopy equivalent to the identity component 
${\rm SO}(p,q)^0$ of ${\rm SO}(p,q)$, so we may also take $G={\rm SO}(p,q)^0$ and denote this by
 $\widehat{{\rm SO}}(p,q)$ for notational consistency.
Note that ${\rm SO}(1)\simeq {\rm pt}$ and $\widehat{{\rm SO}}(1,n)\simeq {\rm SO}(n)$ so that we do not have to consider the case when $n=1$ in killing its higher homotopy groups. 

\medskip
On the other hand, we have the following.

\begin{definition}
The twisted covering, denoted by $\widetilde{SO}({p,q})$,  is the pullback:
\(\xymatrix{
B\widetilde{SO}(p,q) \ar[d] \ar[r] ~& B{\rm O}(q) \ar[d]^-{w_1} \\
B{\rm O}(p)\ar[r]^-{w_1} & K(\Z/2,1).
}\)
\end{definition}

\subsection{Indefinite Spin groups}
\label{Sec Spin}

The next step in going from ${\rm SO}(n)$ to ${\rm Spin}(n)$ by taking a double cover, or killing the fundamental group,
leads to
the isomorphism
\(
\pi_i({\rm Spin}(n))=\begin{cases}
0 & \text{ for }i\geq 2\\
\pi_i({\rm O}(n))&\text{ for }i\geq 3.
\end{cases}
\)
Note the following useful isomorphisms in low degrees:
${\rm Spin}(2) \cong {\rm U}(1)$, 
${\rm Spin}(3) \cong {\rm SU}(2)$,
 ${\rm Spin}(5)\cong {\ Sp}(2)$, and ${\rm Spin}(6)\cong{\rm SU}(4)$.

 \medskip
 We next consider the indefinite orthogonal group ${\rm SO}(p, q)$ and kill the first homotopy 
 group  $\pi_1$. Here there are two case, depending on whether one of the factors $p$ or $q$ 
 is greater than 1. Thus we would like to kill the fundamental group of 
 either ${\rm SO}(1,n)\simeq {\rm O}(n)\langle 1\rangle\simeq {\rm SO}(n)$ 
 or ${\rm SO}(p,q)\simeq {\rm O}(p,q)\langle 1\rangle$ for $p,q\geq 2$.
 An important distinction with the definite case is that the maximal 
 compact subgroup of ${\rm SO}(p, q)$ is ${\rm SO}(p) \times {\rm SO}(q)$, signalling 
 that connectedness involves more than just the usual $\Z/2$.

\begin{remark} 
{\bf (i).} The group ${\rm Spin}(p, q)$ is defined to be the double cover of 
${\rm SO}(p, q)$ in a slightly nontrivial way, namely the cover corresponding to the 
diagonal $\Z/2$ inside $\Z/2 \times \Z/2$. This diagonal $\Z/2$ is formed of a pair of elements,
 where the first correspond to the kernel of ${\rm Spin}(p) \to {\rm SO}(p)$ and the second to 
 ${\rm Spin}(q) \to {\rm SO}(q)$. 
  For example, ${\rm Spin}(2, 2) \cong {\rm SL}(2, \R) \times {\rm SL}(2, \R)$.  
  See \cite{Wo} \cite{Ba} \cite{LM} \cite{Tr} \cite{Va}. 
 
 \vspace{2mm}
\noindent {\bf (ii).} The condition for having  a ${\rm Spin}(p, q)$ structure from a ${\rm SO}(p, q)$ 
structure is the  separate vanishing of two second Stiefel-Whitney classes $w_2^i$, $i=1, 2$ 
(see \cite{Sh} for details).  
 \end{remark}

 \medskip
 Rationally, the cohomology ring of the special orthogonal group is given as
 \(
H^*(B{\rm SO}(n); \Q) \cong 
\begin{cases}
\Q\big[ p_1, p_2, \cdots, p_{[\tfrac{n}{2}]}\big] , & n~{\rm odd},
\\
\Q\big[ p_1, p_2, \cdots, p_{\tfrac{n}{2}}, e\big]/\big(p_{\tfrac{n}{2}}-e^2\big) , & n~{\rm even},
\end{cases}
 \)
 where $p_i$ are the Pontrjagin classes in degree $4i$ and $e$ is the Euler class in 
 degree $n$. The result on the odd case is what one expects, while the even case introduces 
 a new generator. When considering the integral case, this generator persists and, in addition, 
 we will have other generators arising from integral lifts of the Steifel-Whitney classes, i.e.
 arising from applying the Bockstein on monomials in even Stiefel-Whitney classes. Since we 
 are interested in degree four generators, the latter will not be of relevance to us.

\medskip
We know that $H^2(B{\rm SO}(n);\Z/2)\cong \Z/2$ with the second Stiefel-Whitney class $w_2$ as its generator. 
For $n=2$, the integral cohomology of $B{\rm SO}(2)$ is isomorphic to $\Z$ with a single generator 
\footnote{Here we are calling the generators $\sqrt{p_1}$ as these square to the first Pontrjagin class. 
At the level of $\Z/2$ coefficients, this is reminiscent of relations such as $w_2^2 \equiv p_1$ mod 2. In fact, this generator is 
the Euler class.}
$\sqrt{p_{1}}$ 
such that $\sqrt{p_{1}}\sqrt{p_{1}}=p_1\in H^4(B{\rm SO}(2);\Z)$ by the result of Brown
on the integral cohomology ring of $B{\rm SO}(n)$ \cite{Br}. In general, in the integral cohomology of 
$B{\rm SO}(n)$ the square $e^2$ of the Euler class is the same as the Pontrjagin class
in degree $4n$. So for $n=1$, we have a generator of degree 2 given by $e=\sqrt{p_1}$. 
One can also view this as a first Chern class if one identifies ${\rm SO}(2)$ with ${\rm U}(1)$
and hence $B{\rm SO}(n)$ with $\C P^\infty$, whose cohomology is given as
$H^*(\C P^\infty; \Z)\cong \Z[x]$, with $|x|=2$.

\medskip
Therefore, we obtain a $1$-connected cover $G(n)$  of ${\rm SO}(n)$ for $n\geq 3$, and $G(2)$ of
${\rm SO}(2)$, respectively, by taking pullbacks:
\(
\xymatrix{
& BG(n)\ar[d]\ar[r]& {\rm pt}\ar[d]\\
X\ar[r] \ar[ur]& B{\rm SO}(n)\ar[r]_-{w_2} & K(\Z/2,2)\;,
}\qquad \qquad \xymatrix{
& BG(2)\ar[d]\ar[r]& {\rm pt}\ar[d]\\
X\ar[r] \ar[ur]& B{\rm SO}(2)\ar[r]_-{\sqrt{p_1}} & K(\Z,2)\;.
}
\)
We know that the spin group ${\rm Spin}(n)$ is homotopy equivalent to $G(n)$ for $n\geq 3$.
However, ${\rm Spin}(2)$ and $G(2)$ do not agree since ${\rm Spin}(2)\cong S^1$ is not simply connected. 
\begin{definition} 
The \emph{homotopic (definite) spin group}, denoted by $\widehat{{\rm Spin}}(n)$, is the $1$-connected cover of ${\rm SO}(n)$ for all $n\geq 2$ so that $\widehat{{\rm Spin}}(n)\simeq {\rm Spin}(n)$ and $\widehat{{\rm Spin}}(2)\simeq G(2)$. 
\end{definition}
In fact, as far as killing higher homotopy groups is concerned, $G(2)$ is trivial and we do  not have to consider the case with $n=2$ in the process.

\medskip
Next, in order to kill $\pi_1$ of ${\rm SO}({p,q})$, we consider three cases: When $p=q=2$, when 
either $p$ or $q$ is equal to 2, and when both $p$ and $q$ are greater than 2. 
\begin{definition} 
The Spin groups in the above three cases will be the pullbacks in the following diagrams
for $p,q\geq 3$:
$$\xymatrix{
BG(2,2) \ar[rr] \ar[d] && {\rm pt}\ar[d]\\
B{\rm SO}(2,2) \ar[rr]_-{\sqrt{p_1}\times \sqrt{p_1}}&& K(\Z\times\Z,2),
} \xymatrix{
BG(2,q) \ar[r] \ar[d] & {\rm pt}\ar[d]\\
B{\rm SO}(2,q) \ar[r]_-{\sqrt{p_1}\times w_2}& K(\Z\times\Z/2,2),
} \xymatrix{
BG(p,q) \ar[r] \ar[d] & {\rm pt}\ar[d]\\
B{\rm SO}(p,q) \ar[r]_-{w_2\times w_2}& K(\Z/2\times\Z/2,2).
}
$$
\end{definition} 
To justify this, we need to consider the cohomology groups. For that, we first need 
some calculations.
\begin{lemma} 
\label{HkBSO} The following table gives the homology groups with integral 
coefficient\\$H_k(B{\rm SO}(n);\Z)$ for $k=0,1$ and $2$:
\begin{center}\begin{tabular}{c | c c c}
 & $n=1$ & $n=2$ & $n\geq 3$\\
\hline
$H_0$ &$\Z$ &$\Z$ &$\Z$\\
$H_1$ &$0$ &$0$ &$0$\\
$H_2$ &$0$ &$\Z$& {$\Z/2$}
\end{tabular}\end{center}
\end{lemma}
\proof 
${\rm SO}(1)\cong \{1\}$ and so this case has trivial homology in nonzero degrees. 
Since ${\rm SO}(2)\cong S^1$, we have $B{\rm SO}(2)\cong \C P^\infty$.
We know that $$H_k(\C P^n;\Z)=\begin{cases} \Z, &\text{ if }0\le k\le 2n \text{ and $k$ is even},\\
0,&\text{ otherwise}.\end{cases}$$
so by using the direct limit $H_k(\C P^\infty;\Z)=\varinjlim_n H_k(\C P^n;\Z)$, we obtain
$$
H_k(\C P^\infty;\Z)=\begin{cases} \Z, &\text{ for $k$ even},\\ 0,&\text{ for $k$ odd}.\end{cases}
$$
For $n\geq 3$, we know 
 $$
 \pi_k(B{\rm SO}(n))\cong \pi_{k-1}({\rm SO}(n))=\begin{cases}0, &\text{ if }k=1\\ {\Z/2}, &\text{ if }k=2. \end{cases}
 $$ 
Hence, $H_k(B{\rm SO}(n);\Z)$ follows from the Hurewicz theorem.
\endofproof


In order not to worry about homotopy groups in degree zero, we work with $\widehat{\rm SO}(p, q)$, the 
connected cover of ${\rm SO}(p, q)$.

\begin{proposition} 
\label{PropKunn} 
For $p,q\geq 2$, there is an isomorphism
$$H^2(B\widehat{{\rm SO}}({p,q});\Z)\cong H^2(B{\rm SO}(p);\Z)\times H^2(B{\rm SO}(q);\Z).$$
\end{proposition}

\proof 
For any positive integer $p$ and $q$, the K\"unneth formula gives
\begin{align*}
H^2(B{\rm SO}(p)\times B{\rm SO}(q);\Z)&\cong \hom(H_2(B{\rm SO}(p)\times B{\rm SO}(q);\Z),\Z)\\
&\ \ \oplus {\rm Ext}_\Z^1(H_1(B{\rm SO}(p)\times B{\rm SO}(q);\Z),\Z)\;.
\end{align*}
Now the homology groups inside the hom and Ext factors on the right hand side are calculated as
\begin{align*}
H_2(& B{\rm SO}(p)\times B{\rm SO}(q);\Z)\cong (\bigoplus_{r+s=2}H_r(B{\rm SO}(p);\Z)\otimes_\Z H_s(B{\rm SO}(q);\Z))\\
& \qquad \qquad \qquad \qquad \qquad \
\oplus(\bigoplus_{r+s=1}{\rm Tor}_1^\Z(H_r(B{\rm SO}(p);\Z),H_s(B{\rm SO}(q);\Z)))\\
&\cong\begin{cases}
0, &\text{ if $p=1,q>2$ or $p>2, q=1$},\\
\Z, &\text{ if $p=1,q=2$ or $p=2,q=1$},\\
 \Z\oplus \Z\cong H_2(B{\rm SO}(p);\Z)\oplus H_2(B{\rm SO}(p);\Z), &\text{ if $p,q = 2$}\\
{ \Z/2\oplus \Z/2}\cong H_2(B{\rm SO}(p);\Z)\oplus H_2(B{\rm SO}(p);\Z), &\text{ if $p,q \geq 3 $},
\end{cases}
\end{align*}
and
\begin{align*}
H_1(B{\rm SO}(p)\times B{\rm SO}(q);\Z)&\cong (\bigoplus_{r+s=1}H_r(B{\rm SO}(p);\Z)\otimes_\Z H_s(B{\rm SO}(q);\Z))\\
& \ \ \oplus(\bigoplus_{r+s=0}{\rm Tor}_1^\Z(H_r(B{\rm SO}(p);\Z),H_s(B{\rm SO}(q);\Z)))\\
&=0\;,
\end{align*}
since $H_1(B{\rm SO}(n);\Z)$ is trivial from Lemma \ref{HkBSO}. 

\endofproof

%
%

\subsection{Indefinite String groups}
\label{Sec String}

In this section we take as our starting point the indefinite Spin group ${\rm Spin}(p, q)$.
We emphasize that there is a subtlety here in that this group is {\it not} simply connected
for general $p$ and $q$. In fact, the maximal compact subgroup of ${\rm Spin}(p, q)$
is ${\rm Spin}(p) \times {\rm Spin}(q)/\{(1,1), (-1,-1)\}$. The group ${\rm Spin}(p, q)$
itself is the diagonal 2-fold cover of the 4-fold cover of ${\rm SO}(p, q)$. For $p \geq q$,
the fundamental group is given as 
\(
\pi_1({\rm Spin}(p, q))=
\left\{
\begin{array}{llll}
\{0\}  &&& (p, q)=(1,1)~{\rm or}~(1,0)\;,  
\\
\{0\} &&& p>2, q=0,1\;,
\\
\Z &&& (p, q)=(2,0)~{\rm or}~(2,1)\;,
\\
\Z \times \Z &&& (p, q)=(2,2)\;,
\\
\Z &&& p>2, q=2\;,
\\
\Z/2 &&& p, q >2\;.
\end{array}
\right.
\)
In order to 
define indefinite String structure properly we need to take as a starting point a simply-connected
group. Therefore, we should start from the simply-connected cover of ${\rm Spin}(p, q)$, which is 
what we do below.
Note, however, that one can define variants of String structures without requiring this.
The resulting structure would be analogous to the case of $p_1$-structures
(see \cite{S1} \cite{S2} for various analogous structures in the definite case).

\medskip
We next kill the next nontrivial homotopy groups, namely $\pi_3$, of the relevant Spin group ${\rm Spin}(p,q)$. 
Note that at this stage if either $p$ or $q$ is less than 3 then the corresponding factor in the decomposition 
${\rm Spin}(p) \times {\rm Spin}(q)$ will not be seen in the process. So we will consider mainly two cases:
${\rm Spin}(n)\simeq\widehat{{\rm Spin}}(1,n)\simeq \widehat{{\rm Spin}}(2,n)\simeq {\rm O}(n)\langle 3\rangle$ 
and $\widehat{{\rm Spin}}({p,q})\simeq {\rm O}({p,q})\langle 3\rangle$ for $p,q\geq 3$.

\medskip
When $p,q\geq 3$, the maximal compact subgroup of the 1-connected cover ${\rm O}(p,q)\langle 1\rangle$
 is ${\rm O}(p)\langle 1\rangle\times {\rm O}(q)\langle 1\rangle$ which is homotopy equivalent to 
 ${\rm Spin}(p)\times {\rm Spin}(q)$. Hence, ${\rm O}(p,q)\langle 1\rangle$ is homotopy equivalent to ${\rm Spin}(p)\times{\rm Spin}(q)$.
In fact, since ${\rm O}(1)\langle 1\rangle$ and ${\rm O}(2)\langle 1\rangle$ are just a point space and 
contractible space respectively, we still can say that ${\rm O}(p,q)\langle 1\rangle$ is homotopy equivalent to 
${\rm O}(p)\langle 1\rangle\times{\rm O}(q)\langle 1\rangle$.

\medskip
Next we would like to consider cohomology. Here the cohomology groups and rings of the orthogonal and Spin groups 
in the unstable range are quite subtle \cite{Br} \cite{Fe} \cite{Ko} \cite{BR}. However, we will only 
need the degree four group.
Indeed, MacLaughlin has shown \cite{Mc} 
that the fourth cohomology group is $H^4(B{\rm Spin}(n);\Z)\cong \Z$ and is generated by $\frac{1}{2}p_1$. 
Note that this can also be deduced from other means, for instance, 
from the calculations presented by Kono \cite{Ko} and Benson and Wood \cite{BR}.

\medskip
We now get back to the construction of indefinite String structures. 
We obtain a $3$-connected cover $G(n)$ for $n\geq 3$ by taking homotopy
pullbacks, contrasting two cases:
\(
\xymatrix{
BG(n) \ar[r] \ar[d]  & {\rm pt}\ar[d]\\
B{\rm Spin}({n\ne 4})\ar[r]^-{\frac{1}{2}p_1} & K(\Z,4)
}
\qquad \qquad
\xymatrix{
BG(4) \ar[rr] \ar[d]  && {\rm pt}\ar[d]\\
B{\rm Spin}({4})\ar[rr]^-{(\frac{1}{2}p_1, \frac{1}{2}p_1)} && K(\Z\times\Z,4)\;.
}
\)
Here, the map 
$(\frac{1}{2}p_1,\frac{1}{2}p_1):B{\rm Spin}(4)\to K(\Z\times\Z,4)$
 is equivalent to the map 
$\frac{1}{2}p_1\times \frac{1}{2}p_1:B{\rm Spin}(3)\times B{\rm Spin}(3)\to K(\Z\times\Z,4)$ through 
the accidental isomorphism ${\rm Spin}(4)\cong {\rm Spin}(3)\times{\rm Spin}(3)$. Consequently, 
any classifying map 
\footnote{Here a hat on the group indicates that we are taking the simply connected cover.}
$f:X\to B\widehat{{\rm Spin}}(3,3)$ can be decomposed into a pair 
$f=(f_1,f_2)$ with $f_1,f_2:B{\rm Spin}(3)\to K(\Z,4)$. 
Recall also that for any classifying map $f:X\to B\widehat{{\rm Spin}}({p,q})$ can be decomposed
 into $(f_1,f_2):X\to B\widehat{{\rm Spin}}(p)\times B\widehat{{\rm Spin}}(q)$, due to the 
 homotopy equivalence of the target spaces. 
Hence we can use the additive and multiplicative notations interchangeably.

\medskip
In order to kill $\pi_3$ of $\widehat{{\rm Spin}}({p,q})$, we take the pullbacks according to the following procedure.
 
\begin{definition} The String groups  in the indefinite case are defined as the loop spaces of the corresponding 
classifying spaces, which in turn are defined via the following pullbacks:

\noindent {\bf (i)} for $p,q\geq 5$
\(
\xymatrix@C=4em{
B\widehat{{\rm String}}({p,q}) \ar[r] \ar[d] & {\rm pt}\ar[d]\\
B\widehat{{\rm Spin}}({p,q})\ar[r]^-{\frac{1}{2}p_1\times \frac{1}{2}p_1} & K(\Z\times\Z,4)\;;
}
\)
{\bf (ii)} for $p=4, q\geq 5$,
\(
\xymatrix@C=7em{
B\widehat{{\rm String}}({4,q}) \ar[r] \ar[d]  & {\rm pt}\ar[d]\\
B\widehat{{\rm Spin}}({4,q})\ar[r]^-{(\frac{1}{2}p_1\times \frac{1}{2}p_1)\times \frac{1}{2}p_1} & K(\Z\times\Z\times \Z,4)\;;
}
\)
{\bf (iii)} and for 
$p=q=4$, 
\(
\xymatrix@C=9em{
B\widehat{{\rm String}}({4,4}) \ar[r] \ar[d] & {\rm pt}\ar[d]\\
B\widehat{{\rm Spin}}({4,n})\ar[r]^-{(\frac{1}{2}p_1\times \frac{1}{2}p_1)\times (\frac{1}{2}p_1\times \frac{1}{2}p_1)} & K(\Z\times\Z\times \Z\times\Z,4)\;.
}
\)
\end{definition}
Note that we can also consider variants of String structures associated to the non-simply connected
groups ${\rm Spin}(p, q)$. 

\begin{definition}
A $(p_1, p_1')$-structure is a lift from $B{\rm Spin}(p, q)$ to the classifying space obtained by 
killing the fourth homotopy group. 
\end{definition}
These are analogs of a $p_1$-structure, where the lower homotopy groups are not necessarily killed. 
See \cite{S1} \cite{S2} \cite{9brane} for extensions and applications.

\medskip
Of course, we need to establish a decomposition of the corresponding cohomology groups. 
This in turn will require taking homomorphisms
 with homology. 
%
%
%
 To that end, we start with the following:

\begin{lemma} 
$$
H_4(B{\rm Spin}(p)\times B{\rm Spin}(q);\Z) \cong \left\{
\begin{tabular}{ll}
$H_4(B{\rm Spin}(q);\Z)$ & {\rm for~any} $q$  {\rm if} $p=1$, \\

$H_4(B{\rm Spin}(p);\Z)$ & {\rm for any} $p$  {\rm if } $q=1$, \\

$H_4(B{\rm Spin}(p);\Z)\oplus H_4(B{\rm Spin}(q);\Z)$ & {\rm if}~ $p,q\geq 2$.
\end{tabular}
\right.
$$
\end{lemma} 
\proof 
The K\"unneth formula for homology gives the following identity:
\begin{align*}
H_4(B{\rm Spin}(p)\times B{\rm Spin}(q) &;\Z)\cong\big(\bigoplus_{r+s=4}
H_r(B{\rm Spin}(p);\Z)\otimes_\Z H_s(B{\rm Spin}(q);\Z)\big)\\
& \qquad\quad\oplus\big(\bigoplus_{r+s=3}{\rm Tor}_1^\Z(H_r(B{\rm Spin}(p);\Z), H_s(B{\rm Spin}(q);\Z))\big).
\end{align*}
Since $H_s(B{\rm Spin}(q);\Z)=0$ for $s=1,2,3$, the only nontrivial term in ${\rm Tor}$ is $${\rm Tor}_1^\Z(H_3(B{\rm Spin}(p);\Z),H_0(B{\rm Spin}(q);\Z)).$$ 
This is also trivial for $p\geq 2$. Moreover, when $p=1$, we have ${\rm Tor}_1^\Z(\Z/2,\Z)$ and this is trivial since $\Z$ is torsion-free.

The direct sum term on the right hand side of the above 
K\"unneth formula has only two nontrivial factors: 
$H_0(B{\rm Spin}(p);\Z)\otimes H_4(B{\rm Spin}(q);\Z)$ and $H_4(B{\rm Spin}(p);\Z)\otimes H_0(B{\rm Spin}(q);\Z)$. 
When $p\geq 2$, these two are isomorphic to $\Z\oplus\Z\cong \Z$. 
At this stage, there seems to be several routes to take.
We have the isomorphisms
\begin{align*}
H_0(B{\rm Spin}(p);\Z)\otimes H_4(B{\rm Spin}(q);\Z)&\cong H_4(B{\rm Spin}(q);\Z),\\
H_4(B{\rm Spin}(p);\Z)\otimes H_0(B{\rm Spin}(q);\Z)&\cong H_4(B{\rm Spin}(p);\Z).
\end{align*}
On the other hand, when $p=1$, we have $H_4(B{\rm Spin}(1);\Z)=0$. So the only nontrivial term is now $H_0(B{\rm Spin}(p);\Z)\otimes H_4(B{\rm Spin}(q);\Z)\cong H_4(B{\rm Spin}(q);\Z)$.
\endofproof

We will also need to calculate the Ext-term.

\begin{lemma} 
$${\rm Ext}_\Z^1(H_3(B{\rm Spin}(p)\times B{\rm Spin}(q);\Z),\Z)\cong 
\begin{cases}\Z/2&{\rm if }~ p=1 {\rm or }~q=1,\\ 0 & {\rm if }~p,q\geq 2.\end{cases}
$$
\end{lemma}
\proof 
First, we need to compute $H_3(B{\rm Spin}(p)\times B{\rm Spin}(q);\Z)$ and we use the K\"unneth formula:
\begin{align*}
H_3(B{\rm Spin}(p)\times B{\rm Spin}(q);&\Z)\cong\big(\bigoplus_{r+s=3}
H_r(B{\rm Spin}(p);\Z)\otimes_\Z H_s(B{\rm Spin}(q);\Z)\big)\\
&\qquad\oplus\big(\bigoplus_{r+s=2}{\rm Tor}_1^\Z(H_r(B{\rm Spin}(p);\Z), H_s(B{\rm Spin}(q);\Z))\big).
\end{align*}
The ${\rm Tor}$ term is trivial since $H_s(B{\rm Spin}(q);\Z)=0$ or $\Z$, and $\Z$ is torsion-free.
The only nontrivial factor in the first term on the right hand side is $H_3(B{\rm Spin}(p);\Z)\otimes H_0(B{\rm Spin}(q);\Z)$. This is zero for $p\geq 2$ since $H_3(B{\rm Spin}(p);\Z)=0$. On the other hand, if $p=1$, we have $H_3(B{\rm Spin}(1);\Z)=\Z/2$ and the result follows from the relation $\Z/2\otimes \Z\cong \Z/2$.
\endofproof

We are now ready to calculate the degree four cohomology groups.

\begin{proposition} 
\label{H4BSpin} 
For any $p, q \geq 2$,
we have
$$H^4(B{\rm Spin}(p)\times B{\rm Spin}(q);\Z)\cong H^4(B{\rm Spin}(p);\Z)\oplus H^4(B{\rm Spin}(q);\Z)\;.$$
For the lower dimensional case, we have 
$$H^4(B{\rm Spin}(1)\times B{\rm Spin}(q);\Z)\cong 
H^4(B{\rm Spin}(q);\Z)\;.$$
\end{proposition} 
\proof 
The K\"unneth formula for cohomology asserts that
\begin{align*}
H^4(B{\rm Spin}(p)\times B{\rm Spin}(q);\Z)&\cong \hom (H_4(B{\rm Spin}(p)\times B{\rm Spin}(q);\Z),\Z)\\
& \ \ \oplus {\rm Ext}_\Z^1(H_3(B{\rm Spin}(p)\times B{\rm Spin}(q);\Z),\Z)\;.
\end{align*}
When $p=1$, by the above lemmas, we have 
\begin{align*}
H^4(B{\rm Spin}(1)\times B{\rm Spin}(q);\Z)&\cong \hom(H_4(B{\rm Spin}(q);\Z)\oplus \Z/2\\
& \cong H^4(B{\rm Spin}(1);\Z)\times H^4(B{\rm Spin}(q);\Z)\;.
\end{align*}
Here, we used the fact that finite products and finite coproducts coincide in the additive category.
When $p\geq 2$, we have
\begin{align*}
H^4(B{\rm Spin}(p)\times B{\rm Spin}(q);\Z) & \cong \hom(H_4(B{\rm Spin}(p);\Z)\oplus H_4(B{\rm Spin}(q);\Z),\Z)\\
& \cong \hom(H_4(B{\rm Spin}(p);\Z),\Z)\times \hom(H_4(B{\rm Spin}(q);\Z),\Z)\\
&\cong H^4(B{\rm Spin}(p);\Z)\times H^4(B{\rm Spin}(q);\Z)\;.
\end{align*}
For $p=2$, we have $H_4(B{\rm Spin}(2); \Z)\cong H_4(\CP^\infty; \Z) \cong \Z$, so that 
$H^4(B{\rm Spin}(2); \Z)\cong \Z$. 

The maximal compact subgroup of $G={\rm Spin}(p,q)$ is $K={\rm Spin}(p)\times{\rm Spin}(q)/(\Z/2)$. 
Therefore, $G$ and $K$ are weakly homotopy equivalent to each other so that they are in fact homotopy 
equivalent, since the usual cohomology is represented  by the Eilenberg-MacLane spaces in a sense that 
$H^n(X;A)\cong [X,K(A,n)]$, where $A$ is a coefficient group (or an integer ring)  and $X$ is arbitrary
 topological space. From the following short exact sequence 
 $$
 1\longrightarrow \Z/2\longrightarrow {\rm Spin}(p)\times{\rm Spin}(q)\longrightarrow ({\rm Spin}(p)\times{\rm Spin}(q))/(\Z/2)\longrightarrow 1
 $$ 
and the fact that $H^n(\Z/2;\Z)=0$ for any $n$, we obtain the desired isomorphism 
$H^4(B{\rm Spin}(p,q);\Z)\cong H^4(B{\rm Spin}(p)\times B{\rm Spin}(q);\Z)$.
\endofproof

\subsection{String structure associated to indefinite unitary and symplectic groups}

\label{Sec U Sp}

\paragraph{The indefinite unitary group.} 

Let $U(p, q)$ denote the group of matrices of linear isometries of the pseudo-Hermitian 
space $\C^{p, q}$ of signature $p, q$. 
The special indefinite unitary group $SU(p, q)={\rm U}(p, q) \cap {\rm SL}(p+q, \C)$ is the subgroup 
of $U(p, q)$ consisting of matrices of determinant 1.

\medskip
The following table lists the homotopy groups of the unitary group \cite{Ker} 
(see \cite{Lu} \cite{PR} for explicit generators).

\begin{center}
\begin{tabular}{|c|cccccc|}
\hline
 & $\quad {\rm U}(1)\quad$ & $\quad {\rm U}(2)\quad$ & $\quad {\rm U}(3)\quad$ & 
$\quad {\rm U}(4)\quad$ & $\quad {\rm U}(5)\quad$ & $\quad {\rm U}(6)\quad$\\
\hline
$\pi_1$ & $\boxed{\Z}$ & $\Z$ & $\Z$ & $\Z$ & $\Z$ & $\Z$\\
$\pi_2$ & $0$ & $0$ & $0$ & $0$ & $0$ & $0$\\
$\pi_3$ & $0$ & $\boxed{\Z}$ & $\Z$ & $\Z$ & $\Z$ & $\Z$\\
$\pi_4$ & $0$ & $\Z/2$ & $0$ & $0$ & $0$ & $0$\\
$\pi_5$ & $0$ & $\Z/2$ & $\boxed{\Z}$ & $\Z$ & $\Z$ & $\Z$\\
$\pi_6$ & $0$ & $\Z/{12}$ & $\Z/{6}$ & {$0$} & $0$ & $0$\\
$\pi_7$ & $0$ & $\Z/2$ & $0$ & $\boxed{\Z}$ & $\Z$ & $\Z$\\
\hline
\end{tabular}
\end{center}

\begin{remark} {\bf (i)} The indefinite unitary group admits a Cartan decomposition 
${\rm U}(p, q) \cong {\rm U}(p) \times {\rm U}(q) \times \C^{pq}$, so that -- as in the case of the orthogonal 
group -- the cohomology is determined by the maximal compact subgroup $K={\rm U}(p) \times {\rm U}(q)$. 
This gives rise to two cohomology classes, one from each factor in $K$, except when 
$p$ or $q$ is equal to 1, in which case there is only one class in real degree four, namely 
$c_2$ for the complementary nontrivial factor in ${\rm U}(1, q)$ or ${\rm U}(p, 1)$.

\vspace{1mm}
\noindent {\bf (ii)} All the unitary groups have nontrivial fundamental group, isomorphic to 
$\Z$. The universal covering groups of the indefinite unitary group and the special indefinite unitary group
are denoted $\widetilde{\rm U}(p, q)$ and 
$\widetilde{\rm SU}(p, q)$,
respectively. Note that the latter is also a subgroup of the former. 

\vspace{1mm}
\noindent {\bf (iii)} As far as $\pi_3$ is concerned, the groups are already in the stable range. 
This makes the discussion much simpler than in the orthogonal case. 

\vspace{1mm}
\noindent {\bf (iv)} The cohomology rings of classifying spaces of 
the unitary group and the special unitary group with 
integral coefficients are generated by the Chern classes $c_i$ in degree $2i$ and has a considerably
simpler form than the orthogonal case, i.e.
\bea
H^*(B{\rm U}(n); \Z) &\cong& \Z[ c_1, c_2, c_3, \cdots , c_n]\;,
\nonumber\\
H^*(B{\rm SU}(n); \Z) &\cong& \Z[c_2, c_3, \cdots , c_n]\;.
\eea
\end{remark}

 String structures associated to the unitary group are considered in constructions associated to 
 elliptic cohomology, such as in \cite{AHS}. Similarly, we have:

\begin{definition} 
A String structure on a space $X$ with an indefinite unitary structure, i.e. with a map $f: X \to  B\widehat{\rm U}(p, q)$ 
is defined via  the following lifting diagram to a map $\tilde{f}$

\noindent {\bf (i)} for $p,q\geq 2$
$$
\xymatrix@C=4em{
&& B\widehat{{\rm String}}({\rm U}({p,q})) \ar[d]  & \\
X \ar[rr]^-f \ar@{-->}[urr]^-{\tilde{f}} && B\widehat{{\rm U}}({p,q})\simeq 
B\widehat{{\rm U}}({p})\times B\widehat{{\rm U}}({q})
\ar[r]^-{c_2 \times c_2'} & K(\Z\times\Z,4)\simeq K(\Z, 4) \times K(\Z, 4)\;,
}
$$
where $c_2, c_2': (E, E') \mapsto (c_2(E), c_2'(E'))$ are the representatives in degree 4 of each of the factor maps;

\noindent {\bf (ii)} for $p=1$ 
$$
\xymatrix@C=4em{
&& B\widehat{{\rm String}}({\rm U}({1,q})) \ar[d]  & \\
X \ar[rr]^-f \ar@{-->}[urr]^-{\tilde{f}} && B\widehat{{\rm U}}({1,q})\simeq 
\times B\widehat{{\rm U}}({q})
\ar[r]^-{ c_2'} & K(\Z, 4)\;,
}
$$
since $B\widehat{{\rm U}}({1})\simeq \ast$.
%
Similarly for $q=1$ with a class $c_2$ corresponding to the first factor.

\end{definition}

The following is immediate from the definition.

\begin{proposition} 
{\bf (i)} ${\rm String}({\rm U}(p, q))$ and ${\rm String}({\rm SU}(p, q))$ structures are classified by a pair of 
classes $(c_2, c_2')$, where $c_2$ and $c_2'$ are the generators in degree 4 of $B{\rm U}(p)$ and $B{\rm U}(q)$, 
respectively. 

\vspace{1mm}
\noindent {\bf (ii)} When either $p=1$ or $q=1$, we only have one generator as an obstruction 
for ${\rm String}({\rm U}(p, q))$.
\end{proposition}

\paragraph{The indefinite symplectic group.} 
The indefinite symplectic group ${\rm Sp}(p, q)$, also known and the indefinite quaternionic unitary group
${\rm U}(p, q; \mathbb{H})$,
can be defined as the isometry group of a nondegenerate quaternionic Hermitian form in $\mathbb{H}^n$.

\begin{remark} {\bf (i)} 
The indefinite symplectic group admits a Cartan decomposition 
${\rm Sp}(p, q) \cong {\rm Sp}(p) \times {\rm Sp}(q) \times \mathbb{H}^{pq}$, so that 
again the cohomology is determined by the maximal compact subgroup $K={\rm Sp}(p) \times {\rm Sp}(q)$,
giving rise to two cohomology classes. Furthermore, due to the relatively large dimension,
there are no degenerate cases here. For instance, ${\rm Sp}(1,1) \cong {\rm Sp}(1) \times {\rm Sp}(1)
\cong S^3 \times S^3$.

\vspace{1mm}
\noindent {\bf (ii)}
The symplectic group is simply connected, so there are no issues with the starting point to 
define a corresponding String structure. 

\vspace{1mm}
\noindent {\bf (iii)} $\pi_i({\rm Sp}(n))$ is in the stable range already for $i \leq 4n+1$. Therefore, 
we are in the stable range for any value of $n$ when considering the third homotopy group.
The homotopy groups 
of the symplectic groups are computed by Mimura and Toda (see \cite{MT}).

\vspace{1mm}
\noindent {\bf (iv)} The cohomology ring of the classifying space of the symplectic group is 
generated by the symplectic Pontrjagin classes $p_i^{\mathbb H}$ of degree $4i$, 
\(
H^*(B{\rm Sp}(n); \Z)\cong \Z[ p_1^\mathbb{H}, p_2^\mathbb{H}, \cdots, p_n^\mathbb{H}]\;.
\)

\vspace{1mm}
\noindent {\bf (v)} Under the identification ${\rm Sp}(1) \cong {\rm SU}(2)$, $p_1^{\mathbb H}$ 
is equal to $-c_2$. 
\end{remark}

\begin{definition} 
A String structure for the indefinite symplectic group ${\rm Sp}(p, q)$ on a space $X$ with 
classifying map $f: X \to B\widehat{{\rm Sp}}({p,q})$ is defined 
as the lifting $\tilde{f}$ in the diagram  
$$
\xymatrix@C=4em{
&& B\widehat{{\rm String}}({\rm Sp}({p,q})) \ar[d]  & \\
X \ar[rr]^-f \ar@{-->}[urr]^-{\tilde{f}} && B\widehat{{\rm Sp}}({p,q})\simeq 
B\widehat{{\rm Sp}}({p})\times B\widehat{{\rm Sp}}({q})
\ar[r]^-{p_1^{\mathbb{H}} \times p_1'^{\mathbb{H}}} & K(\Z\times\Z,4)\simeq K(\Z, 4) \times K(\Z, 4)\;,
}
$$
where $p_1^{\mathbb{H}} \times p_1'^{\mathbb{H}}: (E, E') 
\mapsto (p_1^{\mathbb{H}}(E), p_1'^{\mathbb{H}}(E'))$ are the representatives in
 degree 4 of each of the factor maps.
\end{definition}

As in the unitary case, it follows directly from the definition that we have the following. 

\begin{proposition} 
String structures associated with the indefinite symplectic group are classified by a pair of 
symplectic Pontrjagin classes $(p^{\mathbb{H}}_1, p'^{\mathbb{H}}_1)$, where the first
and second are generators of $H^4(B{\rm Sp}(p); \Z)$ and $H^4(B{\rm Sp}(q); \Z)$,
respectively. 
\end{proposition}

\subsection{Relation to twisted structures}
\label{Sec Twisted}

We have seen that indefinite structures are determined homotopically by their maximal compact
subgroups that are products of two compact Lie groups. The obstruction encountered involves two 
characteristic classes, one from each of these factor group. It is then natural to investigate 
how the two `composite structures' might interact. There is another instance where a pair of
 cohomological structures interact in this context, namely twisted structures 
(\cite{Wang} \cite{SSS3} \cite{S1} \cite{S2} \cite{9brane}), to which we now explore possible 
connections. 

\medskip
We have seen that a $G$-principal bundle $f:X\to BG$ with a topological group $G$ that is homotopy equivalent 
to $G'\times G''$ can be lifted to a $\widehat{G}$-principal bundle, where $\widehat{G}$
  has the same homotopy type as that of $G$ except that $\pi_{n}$ killed, 
 when both obstruction classes $f_1^\ast\alpha'$ and $f_2^\ast \alpha''$ in $H^{n+1}(X;\pi_n(G'\times G''))$ vanish.
  That is, the outer square in the following diagram commutes up to homotopy:
\(
\xymatrix{
X\ar@{-->}[dr] \ar@/^1em/[drr] \ar@/_1em/[ddr]_{f=(f_1,f_2)} &&\\
& B\widetilde{G}\ar[d]\ar[r]& {\rm pt} \ar[d]\\
& B(G'\times G'')\ar[r]^-{\alpha'\times\alpha''} & K(\pi_n(G'\times G''_{n-1}),n+1)\;.
}
\)
Instead of requiring both obstructions $f^\ast_1\alpha'$ and $f_2^\ast \alpha''$ to vanish simultaneously, 
we may want to relax this condition to
just the vanishing of the difference
\begin{equation}
\label{diff eq}
f_1^\ast \alpha'-f^\ast_2\alpha''=0\in H^{n+1}(X;\pi_n(G'\times G'')).
\end{equation}
Now suppose the two groups $G'$ and $G''$ have the same homotopy groups in degree $n$,  i.e. we have
$\pi_{n}(G')\cong \pi_{n}(G'')$.
Denoting this isomorphism group  by $\pi_{n}$,
we take the pullback $\widetilde{G}_0$ as the following diagram:
\(
\xymatrix{
\widetilde{G}_0\ar[r] \ar[d] & BG'' \ar[d]^{\alpha''}\\
BG' \ar[r]^-{\alpha'} & K(\pi_{n},n+1)\;.
}
\)
Suppose that $\widetilde{G}:=\Omega \widetilde{G}_0$ has a topological group structure. 
Given $f_1:X\to BG'$ and $f_2:X\to BG''$ classifying $G'$- and $G''$-principal  bundles over $X$, 
respectively, there is a universal $\widetilde{G}$-principal bundle over $X$ if $f_1^\ast \alpha'$ is homotopic 
to $f_2^\ast\alpha''$ which is equivalent to the condition
\(
f_1^\ast \alpha'-f_2^\ast \alpha'' =0\in H^{n+1}(X;\pi_{n})\;.
\)
Diagrammatically, the condition is equivalent to requiring that there be
 a homotopy $h$ as indicated  in the diagram:
\(
\xymatrix{
X\ar@{-->}[dr]\ar@/_1.5em/[ddr]_{f_1} \ar@/^1.5em/[drr]^{f_2} &&\\
& B\widetilde{G} \ar[r] \ar[d] & BG''\ar[d]^-{\alpha''} \ar@{=>}[dl]_{h} \\
& BG' \ar[r]^-{\alpha'} & K(\pi_{n},n+1)\;.
}
\)
All of this motivates the following definition:
\begin{definition}
 Suppose that we have two topological groups $G'$ and $G''$ with homotopy groups 
$\pi_n':=\pi_n(G')$ and $\pi_n'':=\pi_n(G'')$ respectively.
Moreover, suppose that we have given cohomology classes $\alpha'\in H^{n+1}(BG';\pi_n')$ and 
$\alpha''\in H^{n+1}(BG'';\pi_{n}'')$ and a group homomorphism $\varphi:\pi_n''\to \pi_n'$. Then we 
have the homotopy group $\widetilde{G}$ as in the previous argument, and for two $G'$- and 
$G''$-structures over $X$ given by $f_1$ and $f_2$, the induced $\widetilde{G}$-structure over
 $X$ is said to be \emph{twisted in favor of} $G'$:
\(
\xymatrix{
X\ar@{-->}[dr]\ar@/_1.5em/[ddr]_{f_1} \ar@/^1.5em/[drr]^{f_2} &&\\
& B\widetilde{G} \ar[r] \ar[d] & BG''\ar[d]^-{\varphi\circ \alpha''}  \\
& BG' \ar[r]^-{\alpha'} & K(\pi_{n},n+1)\;.
}
\)
\end{definition}

\begin{remark}
The twisted construction has natural motivations arising from physics, as presented by Sati-Schreiber-Stasheff \cite{SSS3}.
For instance, the Green-Schwarz anomaly condition 
(\cite{GS} \cite{Fr})
\begin{equation}
\label{tw str}
\tfrac{1}{2}p_1(TX)-{\rm ch}_2(E)=0\in H^4(X;\Z)\;,
\end{equation}
where ${\rm ch}_2(E)$ is the second Chern character of a bundle 
$E$ which reduces to the second Chern class $c_2(E)$ is equivalent to existence of a 
homotopy $H_3$ in the following diagram, with $\pi_3({\rm SU}(n))\cong \pi_3({\rm Spin}(n))\cong \Z$ for $n\geq 3$ except $4$,
\(\xymatrix{
X\ar[d]_{TX} \ar[r]^{E} & B{\rm SU}(n)\ar[d]^{c_2} \ar@{=>}[dl]_{H_3} \\
B{\rm Spin}(n)\ar[r]_{\frac{1}{2}p_1} & K(\Z,4)\;.
}\)
The homotopy $H_3$ exhibits the B-field as a \emph{twisted gerbe}, whose twist is the difference class $\frac{1}{2}p_1(TX)-c_2(E)$.  
Our definition above extends this to the indefinite signature case. 
What we have in our current context is what might essentially be viewed as a  twisted 
String structure, in the sense of \cite{Wang} \cite{SSS3}, where the twist itself
arises from a Spin bundle, where the two are the two parts in the 
composite maximal compact subgroup of ${\rm Spin}(p, q)$.
\end{remark}

\begin{remark}
We have only two cases to consider for the twisted coverings:
$$\xymatrix{
B\widetilde{{\rm Spin}}(2,2) \ar[r]\ar[d] & B{\rm SO}(2)\ar[d]^-{\sqrt{p_1}}\\
B{\rm SO}(2)\ar[r]^-{\sqrt{p_1}} & K(\Z,2)
}\qquad {\rm and} \qquad \xymatrix{
B\widetilde{{\rm Spin}}({p,q}) \ar[r]\ar[d] & B{\rm SO}(q)\ar[d]^-{w_2}\\
B{\rm SO}(p)\ar[r]^-{w_2} & K(\Z/2,2)
}
~~~{\rm for}~p,q\geq 3.
$$
\end{remark}

\medskip
Similarly to the previous cases, we can construct twisted coverings:
\begin{definition} 
The twisted covering indefinite String groups are defined as
$$
\xymatrix{
B\widehat{{\rm String}}({p,q}) \ar[r] \ar[d]  & B{\rm String}(q)\ar[d]^{\frac{1}{2}p_1}\\
B{\rm String}(p)\ar[r]^-{\frac{1}{2}p_1} & K(\Z,4)
}\;,
\qquad
\xymatrix@C=4em{
B\widehat{{\rm String}}(4,4) \ar[r] \ar[d]  & B{\rm String}(4)\ar[d]^{\frac{1}{2}p_1\times \frac{1}{2}p_1}\\
B{\rm String}(4)\ar[r]^-{\frac{1}{2}p_1\times\frac{1}{2}p_1} & K(\Z\times\Z,4)\;.
}
$$
\end{definition}

It would be very interesting to extend the definitions and constructions that we  
presented in this paper for String structures to include `pseudo-Riemannian versions' of 
Fivebrane \cite{SSS2} \cite{SSS3} 
and Ninebrane structures \cite{9brane}. This might require considerable calculations. 
We believe that it would also be worthwhile to explore geometric applications 
to gerbes, loop spaces, parallel transport, Chern-Simons theories, and stacky constructions, 
just to name a few. 
We hope to explore these topics elsewhere. Our initial goal was to get to these 
topics directly. However, we realized
 that seemingly straightforward matters are in fact much more subtle 
than meets the eye, so we believe it is worth addressing those first in this paper to 
provide firm ground from which to pursue further constructions.

  \bigskip\bigskip
\noindent
{\bf \large Acknowledgements}
 
\vspace{2mm}
\noindent We would like to thank Domenico Fiorenza, Corbett Redden, Jonathan Rosenberg, and Urs Schreiber for  
very useful discussions.  The research of H. S. was supported by NSF Grant PHY-1102218. We are grateful to  
the anonymous referee for very useful suggestions that have substantially improved the paper.

\end{document}